\documentclass[aps,prb,twocolumn,showpacs,floatfix]{revtex4}
\usepackage{amssymb,amsmath} 
\usepackage{graphicx}
\usepackage{dcolumn}
\usepackage{bm}
\usepackage{color}
\usepackage[colorlinks,bookmarks=false,citecolor=blue,linkcolor=blue,urlcolor=blue]{hyperref}
\def \S{{\bf S}}
\def \R{{\bf R}}
\def \k{{\bf k}}
\def \Q{{\bf Q}}

\def \d{\cdot}
\def\s{\bar{s}}
\def\udel{\underline{\delta}}
\begin{document}
\title{Honeycomb antiferromagnet with a triply degenerate dimer ground state}
\author{Rakesh Kumar}  
\email{rakesh.phys@gmail.com}
\author{Dushyant Kumar}
\author{Brijesh Kumar}
\email{bkumar@mail.jnu.ac.in}
\affiliation{School of Physical Sciences, Jawaharlal Nehru University, New Delhi 110067, India}
\date{\today}
\begin{abstract}
We present an antiferromagnetic quantum spin-1/2 model on honeycomb lattice. It has two parts, one of which is the usual nearest-neighbor Heisenberg model. The other part is a certain multiple spin interaction term, introduced by us, which is exactly solvable for the ground state. Without the Heisenberg part, the model has an exact threefold degenerate dimer ground state. This exact ground state is also noted to exist for the general spin-$S$ case. For the spin-1/2 case, we further carry out the triplon analysis in the ground state, to study the competition between the Heisenberg and the multiple spin interactions. This approximate calculation exhibits a continuous quantum phase transition from the dimer order to N\'eel order.
\end{abstract}
\pacs{75.10.Jm, 75.30.Kz, 75.50.Ee, 75.40.Mg}
\maketitle
\section{Introduction}
\label{sec:introduction}
The low dimensional quantum spin systems are a subject of great current interest. Much of this research directly attends to the real (quasi) one and two dimensional spin systems studied in the laboratories~\cite{kageyama99,ramirez08,moller08,rogado02}. There is also a formal side to it which is concerned with investigating, at various theoretical levels, the effects of low spatial dimensionality, quantum spin fluctuations and frustration on the nature of the ground state of a model spin system. Since the antiferromagnetism is sensitive to all of these, the ground state of a quantum antiferromagnet (AF) can choose from a variety of possibilities (known, or not yet known)~\cite{dagotto_rice96, sachdev08, lee08}. For example, the ground state of the spin-1/2 nearest neighbor (nn) Heisenberg antiferromagnet is a critical spin-liquid (with power law decay of the spin-spin correlations and zero local magnetic moment) in one dimension (1d), but it has N\'eel order on two and higher dimensional bipartite lattices. Furthermore, the competing interactions can induce changes in the nature of ground state, say, from being a N\'eel ordered state to becoming spontaneously dimerized~\cite{kumar08} (or something else). The Majumdar-Ghosh model presents an exactly solvable case of a spontaneously dimerized doubly degenerate singlet ground state in 1d~\cite{majumdar69}. Similar spin models have also been constructed in two dimensions (2d)~\cite{shastry81,klein82,affleck87,gelle08,kumar08}.

Of the spin systems in 2d, the honeycomb lattice comes across as a special case to study. Its site-coordination is three which lies between 1d and the square lattice. Hence, the quantum fluctuations are expected to be stronger on the honeycomb than on the square lattice. Moreover, the honeycomb is not a bravais lattice. It has two spins per unit cell. Thus, we expect a natural case for spontaneous dimer order, without breaking translational symmetry, in a honeycomb antiferromagnet. For the spin-1/2 nn Heisenberg AF on honeycomb lattice, various calculations~\cite{weihong91, reger89, oitmaa92} indeed show larger quantum fluctuations than on the square lattice, but the ground state still exhibits N\'eel order (although weaker than square lattice). Several other studies have shown that under various frustrated conditions the ground state on honeycomb lattice can get disordered~\cite{einarsson91,mattsson94,fouet01,fujimoto05,takano06}. Motivated by these observations, one of us (BK) constructed a quantum spin model with multiple spin interactions on honeycomb lattice, which has an exact triply degenerate dimer ground state (see Fig.~\ref{fig:3fold}).  
Here, we present this model, and investigate using triplon mean-field theory the transition from the dimer to N\'eel order in the ground state, in the presence of the nn Heisenberg AF interaction. This model could be of further interest in investigating deconfined quantum criticality in 2d antiferromagnets~\cite{sachdev08,senthil04,sandvik07}.

The paper is organized as follows. In Sec.~\ref{sec:model}, we discuss the model and its exact ground state. In Sec.~\ref{sec:triplon_mean_field_theory}, we do the triplon mean-field theory. In Sec.~\ref{sec:results_and_discussion}, we discuss the results. Finally, we conclude with a summary.  
%%%%% 
%%%%%
\section{Model}
\label{sec:model}
\begin{figure}[b]
	\centering
		\includegraphics[width=4.5cm]{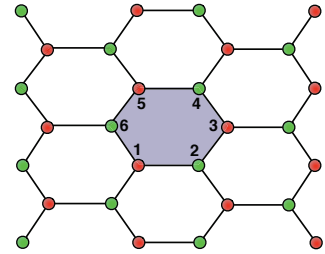}
	\caption{(Color online) The red and green bullets represent the spins on two sublattices of the honeycomb lattice. The connecting lines denote the nearest neighbor antiferromagnetic interaction, $J$. The shaded hexagon denotes the multiple spin interaction, $K$, present on every hexagonal plaquette.}
	\label{fig:model}
\end{figure}

In this paper, we study the following quantum spin-1/2 model on honeycomb lattice (pictorially shown in Fig.~\ref{fig:model}). 
\begin{eqnarray}\label{eq:model}
	  H=J\sum_{\langle ij\rangle} \S_i\d\S_j 
	 + \frac{K}{8}
	\sum_{% 
	\setlength{\unitlength}{0.4cm}
	\begin{picture}(1,1)	
	\put(-.2,.1){\includegraphics[width=.6cm]{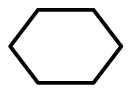}}
	\end{picture}
	}
	\left[\S_{12}^2\S_{34}^2\S_{56}^2
	 +\S_{23}^2\S_{45}^2\S_{61}^2\right]
\end{eqnarray} 
Here,  $\S_{ij}^2=(\S_i+\S_j)^2$. The first term above is the nearest neighbor Heisenberg model with interaction $J$. The second term is what we have introduced to realize the dimer ground state (as in Fig.~\ref{fig:3fold}). This multiple spin interaction is generated by the product of the pairwise total spins of three nn pairs on each hexagon [in two different ways, $(1,2)(3,4)(5,6)$ or $(2,3)(4,5)(6,1)$]. It involves all the six spins of an hexagon, and the summation is taken over all hexagonal plaquettes of the honeycomb lattice with periodic boundary condition. After expanding $\S_{ij}^2$ as $\frac{3}{2}+2 \S_i\d\S_j$ (for a pair of spin-1/2), and regrouping different terms, Eq.~(\ref{eq:model}) becomes:  $H=\frac{27}{64}K L+H^{(2)}+H^{(4)}+H^{(6)}$, where $L$ is the total number of lattice sites, and $H^{(2)}$, $H^{(4)}$ and $H^{(6)}$ denote the quadratic, quartic, and sextic spin interactions, respectively. Note that $H$ respects the lattice translation and point-group symmetries, and is also $SU(2)$ invariant. We take $J, K >0$, and $J+K=1$ sets the unit of energy. Thus, $J=1-K$, where $K\in[0,1]$.
%%%%%
\subsection{Exact ground state}
\label{sub:exact_ground_state}
\begin{figure}[htbp]
	\centering
		\includegraphics[width=4.25cm]{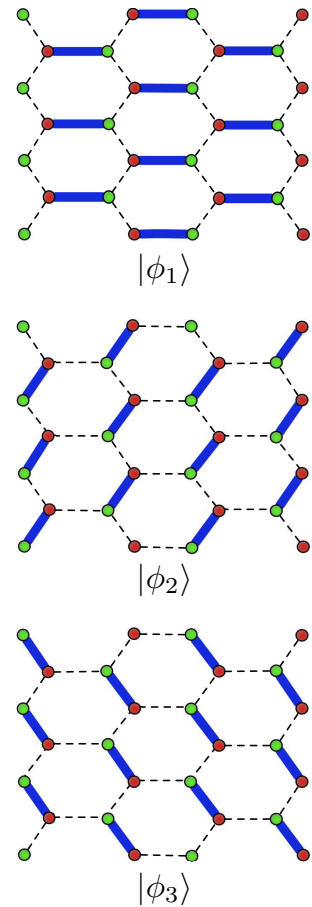}
	\caption{(Color online) The dimer states $|\phi_1\rangle$, $|\phi_2\rangle$, and $|\phi_3\rangle$ form the ground state of the Hamiltonian, $H_K$. The thick blue lines denote the dimer singlets.}
	\label{fig:3fold}
\end{figure}

Consider the model for $K=1$. It can be written as:
\begin{equation}\label{eq:mse}
	H_K=\frac{1}{8}
	\sum_{% 
	\setlength{\unitlength}{0.4cm}
	\begin{picture}(1,1)	
	\put(-.2,.1){\includegraphics[width=.6cm]{honey_little}}
	\end{picture}
	}
	\left[h_{K1}+h_{K2}\right],
\end{equation}
where  $h_{K1}=\S_{12}^2\S_{34}^2\S_{56}^2$ and $h_{K2}=\S_{23}^2\S_{45}^2\S_{61}^2$. In this case, the nn Heisenberg exchange is absent. We only have the multiple spin interactions. Clearly, $h_{K1}$ and $h_{K2}$ have positive eigenvalues, with zero as the minimum. Hence, the ground state energy of $H_K$ is bounded below by zero. The operator $h_{K1}$ gives zero when at least one of three concerned spin pairs, that is (1,2), (3,4) or (5,6), forms a singlet. Similarly for $h_{K2}$. Therefore, on a single hexagon, a zero energy eigenstate of $h_{K1}+h_{K2}$ can be obtained by simultaneously forming singlets on the opposite edges. It leaves the remaining two spins remain as `free'. For example, one such state is $[1,2]\otimes |m_3\rangle\otimes[4,5]\otimes|m_6\rangle$, where $[i,j]=(|\uparrow_i\downarrow_j\rangle - |\downarrow_i\uparrow_j\rangle)/\sqrt{2}$ is the singlet formed by $i$ and $j$ spins, and $m_k=\uparrow$ or $\downarrow$. Moreover, there are three ways of choosing such dimer forming spin pairs. After knowing these single plaquette dimer states, it is  straightforward to show that the three dimer ordered configurations shown in Fig.~\ref{fig:3fold} form the exact zero energy ground state of $H_K$ on the full lattice. We have cross-checked it using numerical diagonalization on a finite spin cluster in the following subsection. These dimer states have been known to arise in the ground state of the quantum dimer model on honeycomb lattice, but ours is probably the first example of a $SU(2)$ spin model on honeycomb with this dimer ground state~\cite{moessner01}.

Let the ground state configurations be denoted as $|\phi_1\rangle$, $|\phi_2\rangle$, and $|\phi_3\rangle$. These dimer states do not break the translational symmetry of the lattice, and are obviously $SU(2)$ invariant. The point group rotational symmetry is broken, however. The wave function of the dimer state, $|\phi_1\rangle$, can be explicitly written as:
\begin{eqnarray}\label{eq:phi1}
	|\phi_1\rangle=\otimes\prod_{(i,j)\in\mathcal{D}}[i,j],
\end{eqnarray}
where $\mathcal{D}$ is the set of singlet forming dimers in the state $|\phi_1\rangle$. The other two states, $|\phi_2\rangle$ and $|\phi_3\rangle$, are related to $|\phi_1\rangle$ via the threefold rotation as:
\begin{subequations}
    \begin{eqnarray}\label{eq:rot_sym}
		|\phi_2\rangle &=& \mathcal{C}_3~|\phi_1\rangle\\
		|\phi_3\rangle &=& \mathcal{C}_3^2~|\phi_1\rangle
	\end{eqnarray}	
\end{subequations}
where $\mathcal{C}_3$ is the clockwise $2\pi/3$ rotation operator. 

To this end, we would like to mention that this exact ground state of $H_K$ is also valid for the general spin-$S$ system. For the spin-$S$ case, a dimer would denote a singlet state formed by a pair of spin-$S$. Everything else (that is, the dimer pattern, the degeneracy, and the ground state energy) is the same. Since the maximum total spin of a pair of spin-$S$ is $2S$, we can rescale the coupling $\frac{K}{8}$ to $\frac{K}{2(2S+1)(2S+1)}$. This just makes the energy contribution of the multiple spin interaction comparable (in powers of $S$) to that of the Heisenberg part.
%%%%%
\subsection{Finite size numerical diagonalization} 
\label{sub:finite_size_diagonalization}
\begin{figure}[htbp]
	\centering
		\includegraphics[width=6.75cm]{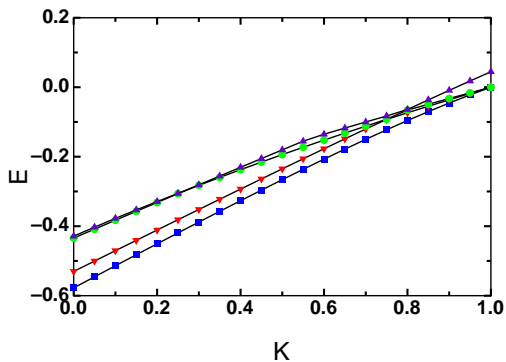}
	\caption{(Color online) Energy eigenvalues per site from the exact diagonalization of $H$ on a 12-site honeycomb cluster.}
	\label{fig:edgs}
\end{figure}

We have done exact numerical diagonalization of $H$ on a $12$-site honeycomb cluster with periodic boundary conditions. This is just to numerically verify the exact ground state on a small cluster. Our present expertise does not allow us to do exact numerical diagonalization on larger spin clusters. We only use total magnetization and spin-inversion symmetries in the coding. The exact diagonalization results clearly show that the ground state for the exactly solvable case ($K=1$) is indeed triply degenerate with zero energy (see Fig.~\ref{fig:edgs}). The next eigenstate has a finite energy gap to the ground state. Away from the exact case,  the ground state energy decreases smoothly without level crossing. Although the degeneracy seems to lift as soon as $K$ is different from $1$, but we expect the degeneracy to survive, in a finite range of $K$ values, for large enough systems.

In order to ascertain the nature of $K=1$ ground state, we compute the spin-spin correlation, $\langle\S_i\d\S_j\rangle$, and  the four-point (dimer-dimer) correlation,
\begin{eqnarray}\label{eq:dimercorr}
	D_{(i,j,k,l)}=\langle(\S_i\d\S_j)(\S_k\d\S_l)\rangle- \langle\S_i\d\S_j\rangle \langle\S_k\d\S_l\rangle.
\end{eqnarray} 
The latter helps in identifying the dimer order. According to this definition, the $D_{i,j,k,l}$ is positive when the two dimers are correlated, and negative if the dimers are uncorrelated. We compute these first in the numerically generated ground state, and then compare them with those calculated in the exactly known dimer ground state. On the infinite lattice, the nn spin-spin correlation is equal to $-1/4$ (further neighbor spin correlations are identically zero in the states of Fig.~\ref{fig:3fold}), and the dimer-dimer correlation is $1/8$ when two dimers are perfect singlets.
\begin{table}
\caption{\label{tab:spincorr}Spin-spin correlations}
\begin{ruledtabular}
\begin{tabular}{lcr}
Dimer & Numerical diagonalization  & Exact\\
\hline
(2,1)  & -0.244768 & -0.244768  \\
(2,3)  & -0.245889 & -0.245889  \\
(2,4)  & -0.008221 & -0.008221  \\
(2,5)  & -0.248879 & -0.248879  \\
(2,6)  & -0.002242 & -0.002242  \\
(2,7)  &  0.000373 &  0.000373  \\
(2,8)  & -0.000373 & -0.000373  \\
(2,9)  &  0.002242 &  0.002242  \\
(2,10) & -0.002242 & -0.002242  \\
(2,11) &  0.000373 &  0.000373  \\
(2,12) & -0.000373 & -0.000373  \\
\end{tabular}
\end{ruledtabular}
\end{table}
                      
For $K=1$, the numerical ground state wavefunctions would be some orthogonal linear combinations of $|\phi_1\rangle$, $|\phi_2\rangle$, and $|\phi_3\rangle$. The choices of the linear combination are not unique, however. As we have not implemented translation and point group symmetries in our computational scheme, this ambiguity in the degenerate output states of our (less sophisticated) program remains. Therefore, we use the zero temperature thermal density operator to correctly compute the ground state properties. For an operator $\hat{\mathcal{O}}$, its thermal average is given by  $\langle \hat{\mathcal{O}}\rangle={\text Tr}(\hat{\rho}\hat{\mathcal{O}})$, 
where $\hat{\rho}=Z^{-1}e^{-\beta\mathcal{H}}$ is the thermal density operator ($Z={\text Tr}~e^{-\beta\mathcal{H}}$). In the zero temperature limit, the density operator reduces to
\begin{eqnarray}\label{eq:dop0k}
	   \hat{\rho}=\frac{1}{N_g}\sum_{\nu=1}^{N_g} |\Psi_{\nu}\rangle \langle\Psi_{\nu}|
\end{eqnarray}
where $N_g$ is the degeneracy of the ground state, and $|\Psi_\nu\rangle$ are the ortho-normalized ground state eigenvectors. In the present calculation, $N_g=3$. 
\begin{table} 	
\caption{\label{tab:dimercorr} Dimer-dimer correlations.}
\begin{ruledtabular}
\begin{tabular}{lcr}
Dimers & Numerical diagonalization & Exact\\
\hline
(2,5)(3,4)  & -0.062599 & -0.062225  \\
(2,5)(4,1)  & -0.061476 & -0.061414  \\
(2,5)(6,7)  & -0.061476 & -0.061414  \\
(2,5)(7,8)  & -0.062599 & -0.062225  \\
(2,5)(9,10) & -0.062599 & -0.062225  \\
(2,5)(10,11)& -0.061476 & -0.061414  \\
(2,5)(11,12)& -0.062599 & -0.062225  \\
(2,5)(12,9) & -0.061476 & -0.061414  \\
(2,5)(4,7)  &  0.122756 &  0.124064  \\
(2,5)(6,9)  &  0.124718 &  0.124718  \\
(2,5)(8,11) &  0.124718 &  0.124718  \\ 
(2,5)(10,1) &  0.124718 &  0.124718  \\
(2,5)(12,3) &  0.124718 &  0.124718  \\
\end{tabular}
\end{ruledtabular}
\end{table}

\begin{figure}[htbp]
	\centering
		\includegraphics[width=5.5cm]{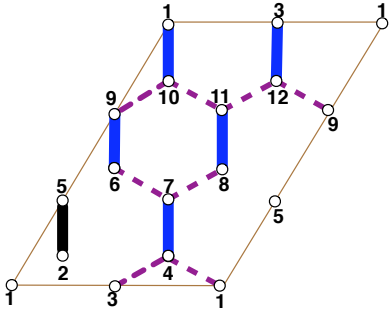}
	\caption{(Color online) The dimer-dimer correlations are calculated by taking the dimer (2,5) as  reference. Thickness of the dimers is proportional to their dimer correlation values. The blue dimer denotes the positive correlation ($i.e.$, a singlet), whereas purple ones represent negative correlation.}
	\label{fig:dimercorr}
\end{figure}

Since the numerical eigenstates are orthonormal, we use them directly to compute the correlations in the ground state, as prescribed above. This data is shown in the second column of the Tables ~\ref{tab:spincorr} and ~\ref{tab:dimercorr}. The exact wavefunctions, $|\phi_1\rangle$ etc. are not orthogonal. Therefore, we first orthogonalize them using Gram-Schmidt procedure (on the same cluster as used for numerical diagonalization; see Fig.~\ref{fig:dimercorr}), then apply the density operator averaging to compute the correlations. These are given in the third column of the two tables. Clearly, the spin correlations are nearest neighbor type, and the dimer correlation matches with dimer order in the exact ground state (compare Fig.~\ref{fig:3fold} with Fig.~\ref{fig:dimercorr}). Numbers from the exact and the numerical calculations match perfectly.  
%%%%%%
%%%%%%
\section{Triplon Mean-Field Theory}
\label{sec:triplon_mean_field_theory}
While at $K=1$, the ground state of $H$ has an exact dimer order, but it is known to be a N\'eel ordered AF state when $K=0$. It would be interesting, therefore, to make some investigation of the transition from the dimer to N\'eel ordered ground state, as $K$ is varied. Here, we present an approximate study of this quantum phase transition by doing triplon analysis with respect to the dimer phase. A {\it triplon} is a triplet excitation residing on a dimer, and dispersing according to the interactions present in the system. While a non-zero gap in the triplon dispersion corresponds to dimer phase, the gaplessness implies AF order in the ground state~\cite{sachdev90,kumar08}.

The triplon analysis is conveniently carried out in the bond operator representation in which the singlet and three triplet states of a pair of spin-1/2 (a bond) are described in terms of the corresponding bosons (called bond operators)~\cite{sachdev90}. The bosonic creation operators, $s^{\dag}$ and $t_{\alpha}^{\dag}$ ($\alpha=x,y,z$), respectively create singlet or triplet states on a bond, subjected to the physical constraint $s^\dag s + t_\alpha^\dag t_\alpha=1$ (repeated Greek indices summed over). The two spins on a dimer are represented as
\begin{subequations}
	\begin{eqnarray}\label{eq:spin2bo}
		S_{1\alpha} &=& \frac{1}{2} \left(s^\dag t_\alpha + t_\alpha^\dag s - 
		i\epsilon_{\alpha\beta\gamma} t_\beta^\dag t_\gamma \right) \\
		S_{2\alpha} &=& \frac{1}{2} \left(-s^\dag t_\alpha - t_\alpha^\dag s - 
		i\epsilon_{\alpha\beta\gamma} t_\beta^\dag t_\gamma \right) 	
	\end{eqnarray}
\end{subequations}
where subscripts $1$ and $2$ denote, say, left and right spins, and $\epsilon_{\alpha\beta\gamma}$ is the totally antisymmetric tensor. In the simplest triplon analysis, the singlet background is treated as a mean field ($s^\dag=\s$), and the triplon dispersion is calculated by ignoring the triplon-triplon interaction (please find a better description of the triplon analysis in Ref.~\onlinecite{kumar08}; we follow the same strategy as therein.).

Although the exact dimer ground state is triply degenerate, we can only take one of these as the reference state to do the triplon mean-field calculation. We rewrite $H$ in terms of the bond operators, taking $|\phi_1\rangle$ as the reference dimer background. The Heisenberg exchange on a dimer can now be written as:
\begin{eqnarray}\label{eq:spinexchange_samed}
	\S_1(\R)\cdot\S_2(\R)=-\frac{3}{4}\s^2+ \frac{1}{4}t^\dag_\alpha t_\alpha,
\end{eqnarray}
where $\R$ is the position vector of the dimer. For the spins coming from different dimers, we have
\begin{widetext}
    \begin{eqnarray}\label{eq:spinexchange_diffd}
		\S_{\mu}(\R)\cdot\S_{\nu}(\R+\udel)=(-1)^{\mu+\nu}~\frac{\s^2}{4}
		\left[t_{\alpha}^{\dag}(\R)t_{\alpha}(\R+\udel)
		+t_{\alpha}(\R)t_{\alpha}(\R+\udel)+h.c.\right]~~~~(\mu,~\nu=1,2).
	\end{eqnarray}	
\end{widetext}
By using Eqs.~(\ref{eq:spinexchange_samed}) and~(\ref{eq:spinexchange_diffd}), and applying the constraint on bond operators globally (with $\mu$ as Lagrange multiplier), we get the following mean-field Hamiltonian for $H$. 
\begin{eqnarray}\label{eq:meanfield}
	H_{mf}&=& E_0 + \frac{1}{2}\sum_{\k} 
	\left\{ \left(\lambda - \s^2\xi_\k\right)
	\left[ t^\dag_{\k\alpha}t^{ }_{\k\alpha} 
	+ t^{ }_{-\k\alpha}t^\dag_{-\k\alpha}\right] \right.\nonumber\\ 
	 &&\hspace{18mm}\left.- \s^2 \xi_\k \left[ t^\dag_{\k\alpha}t^\dag_{-\k\alpha}
	 +t^{ }_{-\k\alpha}t^{ }_{\k\alpha}\right]\right\}
\end{eqnarray}
Here, the triplon operators have been Fourier transformed from the dimer lattice to the corresponding reciprocal lattice, with $\k$ vectors lying in its first Brillouin zone. Moreover, 
\begin{equation}\label{eq:submeanfield}
	E_0=\frac{L}{2}\left[ \frac{J}{4}+\frac{9}{8}K-\left(J+\frac{9}{8}K\right)\s^2
	- \frac{5}{2}\lambda+\lambda\s^2 \right],
\end{equation}
\begin{equation}
	\lambda=\frac{1}{4}\left(J+\frac{9}{8}K\right)-\mu,
\end{equation}
and
\begin{equation}
	\xi_{\k}=\left[ J+\frac{9}{8}K\left(1-\s^2\right)\right] 
	\cos\left(\frac{3}{2}k_x\right) \cos\left(\frac{\sqrt{3}}{2}k_y\right),
\end{equation} 
where $\lambda$ is the effective chemical potential. Equation~(\ref{eq:meanfield}) is diagonalized by the Bogoliubov transformation. The diagonal mean-field Hamiltonian can be written as:
\begin{eqnarray}\label{eq:diaghamilt}
	H_{mf} = E_{0}+\sum_{\k}\, E_{\k}
	\left(\gamma^{\dag}_{\k\alpha}\gamma_{\k\alpha} + \frac{3}{2} \right),
\end{eqnarray} 
where $\gamma_{\k\alpha}$'s are Bogoliubov bosons, and the triplon dispersion, $E_k=\sqrt{\lambda\left(\lambda-2\s^2\xi_k\right)} \ge 0$. The ground state energy per site is given by
\begin{eqnarray}\label{eq:gs_energy}
	e_g[\lambda,\s^2] = e_0+\frac{3}{2L}\sum_{\k}\, E_{\k},
\end{eqnarray} 
where $e_0=E_0/L$. By minimizing the ground state energy with respect to $\lambda$
and $\s^2$, we get the self-consistent equations, whose solution gives the mean-field results.
%%%%%%
\subsection{Dimer phase}
\label{sub:dimer_phase}
When the minimum of the triplon dispersion is nonzero, the dimer phase is stable against triplet excitations. Therefore, the gapped triplon phase corresponds to having dimer ground state.  The self-consistent equations is this case are:
\begin{subequations}
	\label{eq:sce_gapped}
	\begin{eqnarray}
		\s^2&=& \frac{5}{2}-\frac{3}{L}\sum_{\k} \frac{\lambda-\s^2\xi_\k}{E_{\k}} \label{eq:s_gapped}\\
		\lambda &=& J+\frac{9}{8}K + \frac{3\lambda}{L}\sum_\k \frac{\eta_\k}{E_\k} \label{eq:l_gapped}
	\end{eqnarray}	
\end{subequations} 
where $\eta_{\k}=\xi_{\k}-\frac{9}{8}K\s^2\cos\left(\frac{3}{2}k_x\right)
 \cos\left(\frac{\sqrt{3}}{2}k_y\right)$. These equations are obtained by minimizing the ground state energy, i.e., $\partial e_g/\partial\lambda=0$ and $\partial e_g/\partial\s^2=0$.

The weight of having singlet state on a dimer is measured by $\s^2$. If all the dimers form perfect singlets (like in the exact case), then $\s^2=1$. Otherwise, we get $\s^2 <1$, due to triplon fluctuations in the ground state. 
%%%%%%%%%
\subsection{N\'eel phase}
\label{sub:n'eel_phase}
As $K$ is gradually decreased away from $K=1$, at some point we find that the triplon gap vanishes (see Fig.~\ref{fig:dispersion}). The triplon dispersion now touches zero at $\k=\Q$, where $\Q=(0,0)$.  It means the triplon occupancy at wavevector $\Q$ becomes singular, which implies the Bose condensation of triplons at $\Q$. Thus, we need to introduce a third quantity (in addition $\lambda$ and $\s^2$), the triplon condensate density, $n_c$, which is notionally given by
\begin{eqnarray}\label{eq:triplon_density}
	n_c=\frac{2}{L}\langle t^\dag_{\Q\alpha} t_{\Q\alpha}\rangle
	\equiv\frac{3}{L}\left(\frac{\lambda-\s^2\xi_\Q}{E_{\Q}}\right). 
\end{eqnarray} 
Now, the revised set of self-consistent equations are: 
\begin{subequations}
	\label{eq:sce_gapless}
	\begin{eqnarray}
		\lambda &=& 2\s^2\,\xi_\Q \label{eq:l_gapless}\\
		n_c &=& \frac{1}{2\eta_\Q}\left[\lambda-\frac{3\lambda}{L}\sum_{\k\neq \Q}\frac{\eta_\k}{E_\k} -
		\left(J+\frac{9}{8}K\right) \right] \label{eq:nc_gapless}\\
		\s^2&=& \frac{5}{2}-n_c -\frac{3}{L}\sum_{\k\neq \Q} \frac{\lambda-\s^2\xi_\k}{E_{\k}} \label{eq:s_gapless}
	\end{eqnarray}
\end{subequations}
Physically, the non-zero $n_c$ corresponds to AF order in the ground state, which in the present case comes out to be the N\'eel order. The staggered magnetic moment in the N\'eel phase is given by, $M_s=\s\sqrt{n_c}$.
%%%%
%%%%
\section{Results and Discussion}
\label{sec:results_and_discussion}
\begin{figure}[htbp]
	\centering
		\includegraphics[width=6.75cm]{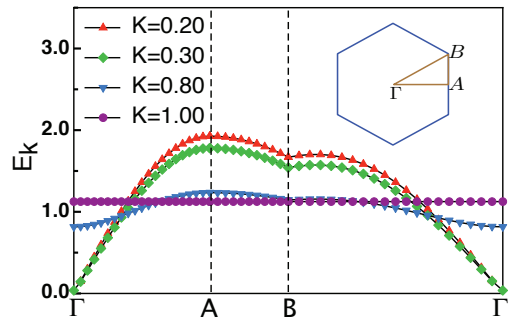}
	\caption{(Color online) Triplon dispersion $E_k$ for different values of $K$. The $\Gamma$, $A$, and $B$ denote the wavevectors $(0,0)$, $(2\pi/3,0)$, and $(2\pi/3, 2\pi/{3\sqrt{3}})$ in the Brillouin zone.}
	\label{fig:dispersion}
\end{figure}

\begin{figure}[htbp]
	\centering
		\includegraphics[width=6.75cm]{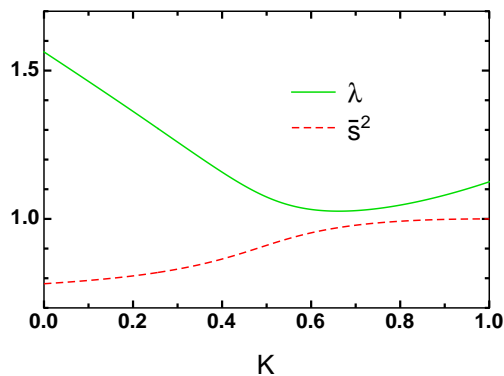}
	\caption{(Color online) The singlet weight, $\s^2$, and the effective chemical potential $\lambda$.}
	\label{fig:lm_sbar}
\end{figure}

Now, we present the results obtained by the triplon mean-field calculation.
The self-consistent Eqs.~(\ref{eq:sce_gapped}) of the gapped phase are solved for $\lambda$ and $\s^2$, for different values of $K$. Interestingly, for $K=1$, it gives $\s^2=1$, same as the exact answer (see Fig.~\ref{fig:lm_sbar}). At the exact point, the triplon dispersion, $E_\k$ (plotted in Fig.~\ref{fig:dispersion}), is flat with an energy gap of $1.125$. 
This value of gap at $K=1$ is in agreement with a direct estimate of $9/8$, which is calculated as: $\Delta_{(K=1)}=\langle\Omega|H_K|\Omega\rangle/\langle\Omega|\Omega\rangle$, where
\begin{eqnarray}
	\label{eq:phi_prim}	 |\Omega\rangle=\left(\otimes\prod_{(i,j)\in\mathcal{D^{\prime}}}[i,j]\right)\otimes\{k,l\}.
\end{eqnarray} 
Here, $\mathcal{D^{\prime}}$ is a set of all dimers in $\mathcal{D}$ except $(k,l)$, and $\{k,l\}$ denotes a triplet state on dimer $(k,l)$. The set $\mathcal{D}$ is same as in $|\phi_1\rangle$ [see Eq.~(\ref{eq:phi1})].

\begin{figure}[htbp]
	\centering
		\includegraphics[width=6.75cm]{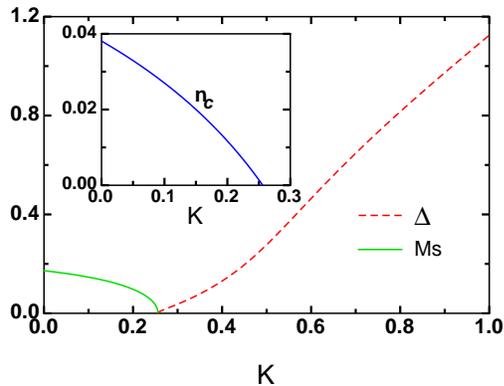}
	\caption{(Color online) The spin gap, $\Delta=E_{\Q}$, and the staggered moment $M_s$. Inset: the triplon condensate density, $n_c$.}
	\label{fig:ms_gap}
\end{figure}

For $K<1$, $E_\k$ acquires a finite width, with minimum at the $\Gamma$ point [that is, $\Q$]. Eventually, it touches zero at $\Q$, and remains so, below $K^*=0.256$. For $K<K^*$, we solve the Eqs.~(\ref{eq:sce_gapless}). As shown in Fig.~\ref{fig:ms_gap}, now the staggered magnetization acquires a non-zero value while the gap remains zero.

\begin{figure}[htbp]
	\centering
		\includegraphics[width=6.75cm]{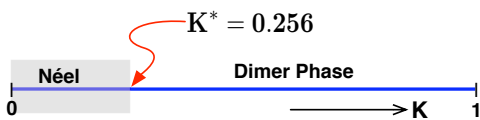}
	\caption{Mean-field quantum phase diagram of $H$ for spin-1/2.}
	\label{fig:qpd}
\end{figure}

The triplon mean-field theory thus predicts a continuous transition from the dimer to N\'eel ordered phase in the ground state of $H$ for spin-1/2. The phase diagram is just a line presented in Fig.~\ref{fig:qpd}. As pointed out earlier, the $K=1$ model has the same exact ground state for higher spins also. Therefore, in some future studies, it would be interesting to extend this quantum phase diagram to include a spin-axis, with $S= 1/2, 1, 3/2, \dots$ to $S\rightarrow\infty$ (the classical limit). The classical case can be discussed right away. Let the spins be classical vectors. The $H_K$ will now have an infinitude of spin configurations in the ground state (not related via global spin rotation), because the spins on each dimer  (of Fig.~\ref{fig:3fold}), separately, must cancel. Hence, $H_K$ itself is a frustrated model. But it does not compete against the classical nn Heisenberg interaction for winning the ground state as the infinite set of pairwise spin-cancelled configurations also includes the N\'eel states. Therefore, $K^*=1$ in the classical limit. This discussion reveals an important feature of  $H$ that is, the multiple spin and Heisenberg interactions don't compete against (or frustrate) each other. Instead, the quantum mechanics acts better when $K$ is sufficiently large. Hence, the N\'eel to dimer transition in the ground state of $H$ is driven purely by quantum fluctuations.

\begin{figure}[htbp]
	\centering
		\includegraphics[width=6.75cm]{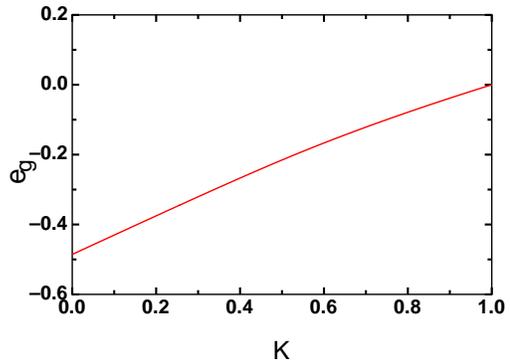}
	\caption{(Color online) The ground state energy per site from the mean-field calculation. It compares reasonably with the exact diagonalization calculation (see Fig.~\ref{fig:edgs}).}
	\label{fig:gsenergy}
\end{figure}

\begin{figure}[htbp]
	\centering
		\includegraphics[width=6.75cm]{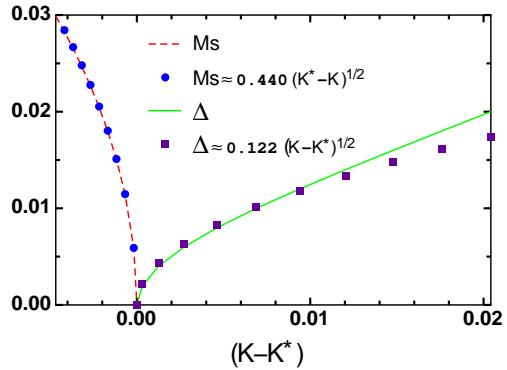}
	\caption{(Color online) The spin-gap and staggered magnetization near the critical point. The estimated results of Eqs.~(\ref{eq:critical_expo}) are also plotted.}
	\label{fig:critical_expo}
\end{figure}

Below, we present the mean-field critical behavior of the spin-gap and staggered magnetization. The critical exponent for both is 1/2, which is derived by analyzing the gap and $n_c$ equations in the small neighborhood of $K^*$.
\begin{subequations}
	\label{eq:critical_expo} 
 	\begin{eqnarray}
		   \Delta&\approx& 0.122 (K-K^*)^{1/2}\\ 
		    M_s&\approx& 0.44(K^*-K)^{1/2}
	\end{eqnarray}   
\end{subequations}
Fig.~\ref{fig:critical_expo} shows an enlarged plot of the mean-field data around $K^*$, together with the estimated critical behavior. The two compare well. 
%%%%%
%%%%%
\section{Summary}
\label{sec:conclusions}
We have constructed and studied a quantum spin-1/2 model on honeycomb lattice. In one limit of the interaction parameter ($K=1$), the model has an exact threefold degenerate dimer ground state. Away from the exact case ($0\le K<1$), we study the evolution of the ground state using triplon mean-field theory. The mean-field theory is exact at $K=1$, and it shows a continuous quantum phase transition from the dimer-ordered to N\'eel ordered ground state at $K^\ast=0.256$. Within this mean-field theory, the critical exponents for the spin-gap (in the dimer phase) and the staggered magnetization (in the N\'eel phase) are 1/2. We have also done preliminary numerical calculations on this model. We have done exact diagonalization on a 12-site honeycomb cluster. For $K=1$, it gives the triply degenerate ground state with the same spin-spin and dimer-dimer correlations as in the exact dimer ground state. We are now focusing on doing numerical work on larger spin clusters.
% section conclusions (end)
%%
\acknowledgments
RK acknowledges CSIR (India) for the scholarship. BK acknowledges the financial support under the project No. SR/FTP/PS-06/2006 from DST (India).
\bibliography{honeycomb}
\end{document}